\documentclass[12pt]{article}

\usepackage{axodraw,cite,graphicx,amssymb}
\usepackage{latexsym,amssymb,epsf,a4wide} 

\newcommand{\beq}{\begin{equation}}
\newcommand{\eeq}{\end{equation}}

\def\Tr{{\rm Tr}}
\def\degree{{\rm deg}}

\def\phibar{\overline{\phi}}

\def\NbyN{N\!\! \times \!\! N}
\def\half{{1 \over 2}}
\def\quarter{{1 \over 4}}
\def\phibar{\overline{\phi}}

\def\Sdef{S_{\rm{def}}}

\def\NbyN{N\! \times\!N}

\begin{document}
\setcounter{page}{0}
\thispagestyle{empty}

\bibliographystyle{hunsrt}

\begin{flushright}
OUTP-00-52P\\
23rd November 2000\\
\end{flushright}
\begin{center}
\vspace{18pt}
{\Large \bf The Cohomological Supercharge}

\vspace{2 truecm}
{\sc Peter Austing}

\vspace{1 truecm}
{\em Department of Physics, University of Oxford \\
Theoretical Physics,\\
1 Keble Road,\\
Oxford OX1 3NP, UK\\ 
\vspace{0.5 truecm}}
p.austing@physics.ox.ac.uk
\vspace{3 truecm}

{\sc Abstract}
\begin{center}
\begin{minipage}{14cm}

We discuss the supersymmetry operator in the cohomological
formulation of dimensionally reduced SYM. By
establishing the cohomology, a large class of 
invariants are classified.
\end{minipage}
\end{center}
\end{center}
\vfill
\begin{flushleft}
PACS: 11.25.-w, 12.60.Jv
\end{flushleft}
\newpage

\section{Introduction}
\setcounter{equation}{0}
A deep relation between dimensionally reduced supersymmetric
Yang-Mills theory, and cohomological field theory was first
uncovered by Moore, Nekrasov and Shatashvili \cite{Moore:1998et}.
To illustrate, we shall consider the four-dimensional Euclidean SYM,
although the techniques which follow can be applied immediately to
the cases of six and ten dimensions. After reduction to zero dimensions, the
action is
\beq
S_{\rm{E}}^{0}= - \Tr \left( \quarter
[X_{\mu},X_{\nu}]^{2} + 
\overline{\lambda} \overline{\sigma}^{\mu}
[X_{\mu},\lambda] \right)
\eeq
All fields are $\NbyN$ 
matrices and transform in the adjoint representation of the gauge
group $G = SU(N)/\mathbb{Z}_{N}$. The
gauge fields
$X_{\mu}$ ($\mu = 1,\cdots, 4$) are restricted to the Lie algebra of
$G$ which is the set of traceless
hermitian matrices. The fermions $\lambda$ are complex traceless
grassman matrices. The $\sigma^{i}$ are the Pauli matrices and
$\sigma^{4}= i \, 1_{2}$. 

A quantity of particular interest is the matrix integral
\beq 
\label{eq:I}
I_{4,N} =
\int dXd\lambda dD \exp  \Tr \left( \quarter
[X_{\mu},X_{\nu}]^{2} + 
\overline{\lambda} \overline{\sigma}^{\mu}
[X_{\mu},\lambda] - 2 D^{2}
\right)
\eeq
(here an auxiliary field $D$ with appropriate integration measure has been
added). In the case of ten dimensions, this integral is of relevance
in its own right as a 
D-instanton partition function \cite{Green:1998yf}, and in a
conjectured non-perturbative description of IIB superstring theory
\cite{Ishibashi:1997xs}. It also arises as the principle part
of the Witten Index in supersymmetric 
Yang-Mills quantum mechanics \cite{Yi:1997eg, Sethi:1997pa}. So it is
of relevance to counting 
threshold bound states of D-particles and, when $N \rightarrow
\infty$, vacuum states in the matrix 
model of M-theory.

Values for $I_{d,N}$ have been calculated analytically for $N=2$
\cite{Yi:1997eg,Sethi:1997pa, Kac:1999av} and numerically for other
small $N$ \cite{Krauth:1998xh,Krauth:1998yu}. In the case of $d=10$,
these results correspond to a conjecture of Green and Gutperle
\cite{Green:1998tn}.

The approach of \cite{Moore:1998et} is to make the following field
replacements.
First rewrite the fermions in terms of their hermitian and antihermitian
parts
\beq
\begin{array}{ccc}
\lambda_{1} & = & (\eta_{2} +i \eta_{1}) \\
\lambda_{2} & = & (\psi_{1} +i \psi_{2}) \\
\end{array}
\eeq
and the auxiliary field
\beq
D = H + \half [X_{1},X_{2}]
\eeq
By the usual contour shifting argument for a gaussian integral, $H$
can be taken hermitian.
To obtain the related cohomological action, make the replacement
\beq
\begin{array}{ccr} 
\phi & = & \half (X_{3} + iX_{4}) \\
\phibar & = & -\half (X_{3} - iX_{4})
\end{array}
\eeq
and take $\phibar$ hermitian and $\phi$ antihermitian. This gives
\beq
\label{eq:Scoh}
\begin{array}{rl}
S_{E}^{0} \rightarrow S_{\rm{coh}} =
\Tr \, {(} & \!\!\!\!
H^{2} + H[X_{1},X_{2}] -\epsilon^{a b} \eta_{1}
[\psi_{a} ,X_{b}]
+\eta_{2} [\psi_{a} ,X_{a} ] \\ & \!\! - \eta_{a} [\phi ,
\eta_{a}] - \psi_{a} 
[\phibar , \psi_{a}] +[X_{a},\phi] [X_{a},\phibar] + [\phi,
\phibar]^{2} \; {)}
\end{array}
\eeq
The key point is that $\phi$ and $\phibar$ are taken independent. Then
$S_{\rm{coh}}$ describes the dimensional reduction of a
Minkowski theory (with lightcone coordinates) whilst $S_{E}^{0}$ gives
the ordinary Euclidean version.

One can easily write the four linearly independent supercharges of
$S_{E}^{0}$ in terms of the new variables, and one of these is
\beq
\label{eq:delta}
\begin{array}{ll}
\delta X_{a} =  \psi_{a} &
\delta \psi_{a}  =  [\phi,X_{a}] \\
\delta \phibar  =  -\eta_{2} & \delta \eta_{2}  =  
-[\phi ,\phibar ] \\
\delta \eta_{1}  =  H & \delta H  =  [\phi , \eta_{1} ] \\
\delta \phi  =  0 
\end{array}
\eeq
Since $\delta^{2}  = [\phi , \;\;]$, $\delta$ is nilpotent on gauge
invariant quantities.

The action is $\delta$-exact. $S= \delta Q$, where
\beq
Q = \Tr (\eta_{1} [X_{1},X_{2}] +  \eta_{1} H -\psi_{a}
[X_{a},\phibar ] - \eta_{2} [\phi , \phibar ] )
\eeq
as can readily be checked. So the symmetry $\delta S =0$ is
manifest. The term cohomological  to describe the
theory is arrived at by analogy of $\delta$ with an exterior
derivative.

In \cite{Moore:1998et}, mass terms are added to the cohomological
action whilst preserving a deformed version of $\delta$. This allows
the cohomological version of the matrix integral 
\beq
\int d\phi dX_{a} d \psi_{a} d\eta_{a} d\phibar dH e^{- \Sdef }
\eeq 
to be calculated.
Finally, the matrix integrals over $\phi$ are replaced with contour
integrals. 

Remarkably, following this prescription gives precisely the same
values as conjectured and obtained numerically for the Yang-Mills
integral (\ref{eq:I}). There is recent evidence that
the prescription also works for some groups other than $SU(N)$
\cite{Staudacher:2000gx,Krauth:2000bv}. An alternative approach also
using the cohomological model is given in \cite{Sugino:1999tv}.

In this note, we study the supersymmetry operator $\delta$, and in
particular we address the question of which quantities are
supersymmetric under $\delta$. Gauge invariant quantities are formed
from traces, and so we seek the general solution to the equation
\beq
\delta \Tr P = 0
\eeq
where $P$ is a polynomial in the matrix fields. In the analogy of
$\delta$ with an exterior derivative, this is the question of finding
the cohomology. An important example of
the use is to find the possible supersymmetric deformations of a
given action. One usually requires a result valid for any gauge group
SU(N), so we shall allow ourselves to make the assumption that $N$ is
suitably 
large. We shall show that
\beq	
\delta \Tr P = 0 \Leftrightarrow \Tr P = \delta \Tr Q + \Tr R( \phi)
\eeq
as long as the  degree of $P$ is less than ${2N \over 3}$.

The proof requires a number of steps. We  form a vector space
from the polynomials, and deal with issues of linear dependence in
section \ref{sec:polys}. A major technical difficulty  is that linearly
independent polynomials become dependent after applying a
trace. This is overcome in section \ref{sec:trace} by forming a suitable
quotient space. Then in section \ref{sec:d} the result is proved for a
simplified version of $\delta$ in which the $[\phi , \;\;]$
terms are absent. Finally, in section \ref{sec:final}, the strands are
drawn together to prove the result.

\section{Cohomology}
\setcounter{equation}{0}
\label{sec:cohom}
\subsection{Polynomials}
\label{sec:polys}

We wish to form a vector space from the polynomials, and eventually
argue by induction on  degree. However, 
there is a technical difficulty. Two polynomials which
\begin{it} look \end{it}
different, because they contain different strings of matrices
multiplied together, can turn out to be identical. At this stage, let
us be definite and make some careful definitions.

\textbf{String} \newline
A string of length $l$ is a map from $\{1,\cdots ,l\}$ into the set of
matrix fields. 
For example, a typical string of length $5$ might be 
\[
\begin{array}{cccccc}
s = & X_{1} & \eta_{2} & \phibar & X_{1} & \phi \\
    & (1)   & (2)      & (3)     & (4)   & (5)
\end{array}
\]

\textbf{Monomial} \newline
A monomial of degree $d>0$ is the matrix product of $d$ matrix fields. The
monomial of degree $0$ is the identity matrix.
For example, a typical monomial of degree $5$ might be
\[
m =  X_{1} \, \cdot \, \eta_{2} \, \cdot \, \phibar \, \cdot \, X_{1}
\, \cdot \, \phi  
\]
where $\, \cdot \,$ indicates matrix multiplication.

Each string of length $l$ is naturally associated to a
monomial of degree $l$ by applying matrix multiplication
between adjacent fields in the string.

\textbf{Polynomial} \newline
A polynomial of degree d is a linear combination of a finite number
of monomials whose highest degree is d.

One can form an abstract vector space $V_{s}$ over $\mathbb{C}$ by taking the
strings as the basis. In $V_{s}$, the strings are linearly
independent. However, as polynomials, the strings are not
necessarily linearly independent. This is most easily seen when the
matrix size $N$ is $1$ so that bosonic matrices commute. Then the two
independent strings $X_{1} \; X_{2}$ and $X_{2} \; X_{1}$ are
identical as polynomials. Even when $N>1$ so that matrices do not
commute, it is possible for independent strings to be linearly
dependent as polynomials. A trivial example is that $\psi^{N^{2}}
\equiv 0$ when $\psi$ is a traceless hermitian $\NbyN$ fermion.

This problem can be overcome by considering only polynomials of degree
smaller than the matrix size. Assume that the matrix fields are $\NbyN$ and
hermitian. They may also have the constraint of tracelessness, but no
other constraints. Then the strings of length less than $N$ are
linearly independent as polynomials.

To see this, 
denote the strings of length less than $N$ by $\{ s^{b} \}$ and the corresponding
monomials $\{ m^{b} \} $.

Suppose 
\beq
\label{eq:ld}
\lambda^{b} m^{b} \equiv 0
\eeq
for some $\lambda^{b} \in
\mathbb{C}$ and (without loss of generality) $\lambda^{1} \neq 0$.

Write $m^{1} = Y^{1} \cdots Y^{d}$ where the $Y^{i}$ are matrix fields
and the degree of $m^{1}$ is $d<N$.

Then, in particular, the term
$Y_{12}^{1} Y_{23}^{2} \cdots Y_{d,d+1}^{d}$ is absent from
(\ref{eq:ld}). But the only monomial which gives rise to this term is
$m^{1} = Y^{1} \cdots Y^{d}$. Therefore $\lambda^{1}=0$, and this is a
contradiction.

Note that  no assumptions are made about which of the matrix fields are
fermionic and which bosonic.

\subsection{The trace}
\label{sec:trace}
It will be convenient  to work with  polynomials rather than traces
of polynomials. Unfortunately, two independent polynomials can have
identical trace. Define an equivalence relation $P \sim Q
\Leftrightarrow \Tr P = \Tr Q$. Then we would like to form the
quotient space $V_{p}/ \! \! \sim$, where $V_{p}$ is the vector space of
polynomials.

Consider a polynomial $P(A^{a})$, where $ \{ A^{a}: a=1,\cdots ,M \} $
are the matrix fields. Assume that the only constraints
which may be applied to the fields are hermiticity and
tracelessness. Define an ordering $\mathcal{O}$
such that
\begin{itemize}
\item $\mathcal{O}$ acts individually on each monomial term in $P$
\item $\mathcal{O}$ cyclically permutes each monomial in $P$ into a
preferred form with a sign to respect fermion statistics
\end{itemize}

An example of such an ordering would be to define
$A^{1}>A^{2}>A^{3}>\cdots$. Then, for example, $\mathcal{O}(A^{4}A^{2}A^{5}) =
(-1)^{F_{4}(F_{2}+F_{5})} A^{2}A^{5}A^{4}$ where $F_{a}$ is the
fermion number of $A^{a}$.

Then, for $\degree (P) <N$
\beq
\label{eq:order}
\Tr P = 0 \Leftrightarrow \mathcal{O}(P) = 0
\eeq
so that $\mathcal{O}$ gives a mapping to the quotient space.

To see that (\ref{eq:order}) is true, first note that
$\mathcal{O} P =0 \Rightarrow \Tr \mathcal{O} P =0 \Rightarrow \Tr P
=0$ by the cyclic property of trace.

Conversely, suppose $\mathcal{O} P \neq 0$. Consider a particular
monomial term in $\mathcal{O} P$:
\[
\mathcal{O} P = \lambda Y^{1} \cdots Y^{M} + \cdots
\]
where  $\lambda$ is some non-zero coefficient.
Then $\Tr P = \Tr \mathcal{O} P$ contains the term:
\[
\Tr P = \lambda Y_{12}^{1} Y_{23}^{2} \cdots Y_{M1}^{M} + \cdots
\]
Since $\mathcal{O} P$ is ordered, the only monomial that can give such
a term is $Y^{1} Y^{2} \cdots Y^{M}$. Therefore this term cannot be
cancelled and so $\Tr P \neq 0$.
Deduce $\Tr P =0 \Rightarrow \mathcal{O} P = 0$. 

An ordering operator $\mathcal{O}$ is not the most useful way of
dealing with the trace. Since there is no way that $\mathcal{O}$
will commute with any form of supersymmetry operator, it is more helpful to
use the following:

Let $P_{M}$ be a polynomial in which all of the terms are of degree
$M$. Then for $M<N$
\beq
\label{eq:cycsum}
\Tr P_{M} = 0 \Leftrightarrow
\sum_{\stackrel{\rm{cyclic}}{\rm{perms}}} P_{M} =0
\eeq

The cyclic permutations act on the matrix fields in the monomials, and
include a sign to respect fermion statistics. For example, suppose $P
= \lambda F_{1} F_{2} B_{3} + \mu B_{1} B_{2} B_{4}$ where the $F_{i}$ are
fermionic and the $B_{i}$ bosonic matrix fields. Then the cyclic
permutation $\sigma 
= (123)$ acts as
\[
\sigma P = - \lambda F_{2} B_{3} F_{1} + \mu B_{2} B_{4} B_{1}
\]

Equation (\ref{eq:cycsum}) follows easily from (\ref{eq:order}).
First note that \[
\sum_{\stackrel{\rm{cyclic}}{\rm{perms}}} P_{M} =
\sum_{\stackrel{\rm{cyclic}}{\rm{perms}}} \mathcal{O} P_{M}
\]
Then
\[
\Tr P_{M} =0 \Rightarrow \mathcal{O} P_{M} = 0 \Rightarrow
\sum_{\stackrel{\rm{cyclic}}{\rm{perms}}} \mathcal{O} P_{M} =0
\Rightarrow \sum_{\stackrel{\rm{cyclic}}{\rm{perms}}} P_{M} =0
\]
Conversely
\[
\Tr \sum_{\stackrel{\rm{cyclic}}{\rm{perms}}} P_{M} = M \Tr P_{M}
\]
and so
\[
\sum_{\stackrel{\rm{cyclic}}{\rm{perms}}} P_{M} = 0 \Rightarrow \Tr
P_{M} = 0
\]

\subsection{Decomposition of the supercharge}
\label{sec:d}
If the expression for $\delta$ (\ref{eq:delta}) did not contain the
commutator terms, our task to classify the supersymmetric quantities
would be much simpler. In this section, we decompose the supercharge
into two parts, and prove our result for the simpler part. This will
allow us to tackle the full supercharge in the next section.
 
Returning to the specific theory under discussion, let us write
\[
A^{1}=X_{1}, \;\; A^{2}=X_{2},\;\; A^{3}=\phibar ,\;\; A^{4}=\eta_{1}
\]
\[
B^{1}=\psi_{1},\;\; B^{2}=\psi_{2},\;\; B^{3}=-\eta_{2} ,\;\; B^{4}=H
\]
Then the supersymmetry $\delta$ can be written as
\beq
\label{eq:genericdelta}
\begin{array}{lcc}
\delta A^{i} = B^{i} & , & \delta B^{i} = [\phi ,A^{i}] \\
\delta \phi = 0
\end{array}
\eeq 
If one considers also the six- and ten-dimensional cohomological matrix
models, one finds an identical form for $\delta$ \cite{Moore:1998et},
and so the results from this point on apply equally to all three theories.

Define two new operators $d$, $\Delta$ by
\beq
\begin{array}{lcc}
d A^{i} = B^{i} & , & d B^{i} = 0 \\
d \phi = 0
\end{array}
\eeq
\beq
\begin{array}{lcc}
\Delta A^{i} = 0 & , & \Delta B^{i} = [\phi ,A^{i}] \\
\Delta \phi = 0
\end{array}
\eeq 
One can very easily check that the following relations hold on
polynomials

\begin{tabular}{rl}
i) & \hspace{1cm} $\delta^{2} = [\phi ,\;\;]$ \\
ii) &\hspace{1cm}  $d^{2} = 0$ \\
iii) & \hspace{1cm} $\Delta^{2} = 0$ \\
iv) & \hspace{1cm} $\delta = d + \Delta$ \\
v) & \hspace{1cm} $\{ d,\Delta \} = \delta^{2} = [\phi ,\;\;]$
\end{tabular}

and that the operator $d$ has the useful property
\beq
\label{eq:dcomm}
d \sum_{\stackrel{\rm{cyclic}}{\rm{perms}}} P_{M} =
\sum_{\stackrel{\rm{cyclic}}{\rm{perms}}} d P_{M}
\eeq
which will allow us to deal with the
trace.

The first step is to prove the result for a polynomial. Let $P$ be a
polynomial of degree less than $N$, and suppose $dP=0$. Then  

\beq
\label{eq:dpoly}
d P=0 \Rightarrow P=dQ+R(\phi )
\eeq 
for some polynomials $Q$ and $R$.

To show (\ref{eq:dpoly}), use
induction on the degree of $P$.

The case $\degree (P)=0$ is simple since then $P = \lambda I = R(\phi )$.

When $\degree (P)>0$, expand
\beq
\label{eq:d0}
P = A^{i} S^{i} + B^{i} T^{i} + \phi U + \lambda I
\eeq
for some polynomials $S^{i}$, $T^{i}$ and $U$, and a constant $\lambda$. 
Apply $d$:
\beq
0 = d P = B^{i} S^{i} + (-1)^{A^{i}} A^{i} d S^{i} + (-1)^{B^{i}}
B^{i} d T^{i} + \phi d U
\eeq
The notation $(-1)^{A^{i}}$ is shorthand for $\pm 1$ respectively as
$A^{i}$ is bosonic or fermionic. Then in particular, since $d$ maps
bosons to fermions, $(-1)^{B^{i}} = (-1)^{A^{i}+1}$.

Since $\degree (P)<N$, the strings are linearly independent (section
\ref{sec:polys}). Deduce 
\beq
\label{eq:d1}
S^{i} + (-1)^{A^{i}+1}d T^{i} = 0
\eeq
\beq
\label{eq:d2}
d S^{i} = 0
\eeq
\beq
\label{eq:d3}
d U = 0
\eeq
Note since $d^{2}=0$, (\ref{eq:d2}) is implied by (\ref{eq:d1}). 

By induction, $dU=0$ implies
\beq
\label{eq:d4}
U = d V + W(\phi )
\eeq
where $V$ and $W$ are polynomials.

Substituting (\ref{eq:d1}) and (\ref{eq:d4}) into (\ref{eq:d0}),
\beq
\begin{array}{ccl}
P &=& A^{i} (-1)^{A^{i}} d T^{i} + B^{i} T^{i} + \phi \left( dV +
W(\phi) \right) +\lambda I \\
&=& d(A^{i} T^{i} + \phi V ) + \phi W(\phi ) + \lambda I \\
&=& d Q + R(\phi)
\end{array}
\eeq
and the result (\ref{eq:dpoly}) follows.

Finally in this section, we introduce the trace.
Let $P$ be a polynomial of degree less than $N$, and suppose $d \Tr P
= 0$. Then 
\beq
\label{eq:dtrace}
d \Tr P = 0 \Rightarrow \Tr P = d \Tr Q + \Tr R(\phi )
\eeq
for some polynomials $Q$
and $R$.

To see this, write $P=P_{0} + \cdots + P_{M}$ where each $P_{i}$
contains only monomials 
of degree $i$.

Then, since $d$ preserves the degree of monomials,
\[
\begin{array}{ccl}
d \Tr P = 0 &  \Rightarrow & d \Tr P_{i} = 0 \\
            &  \Rightarrow & \Tr d P_{i} = 0 \;\;\;\;(i=0,\cdots,M)
\end{array}
\]

The case of $i=0$ is simple. For $i>0$,  using (\ref{eq:cycsum}) gives
\[
\sum_{\stackrel{\rm{cyclic}}{\rm{perms}}} d P_{i} =0
\]
and (\ref{eq:dcomm}) implies
\[
d \sum_{\stackrel{\rm{cyclic}}{\rm{perms}}} P_{i} =0
\]
Then (\ref{eq:dpoly}) gives
\[
\sum_{\stackrel{\rm{cyclic}}{\rm{perms}}} P_{i} = d Q_{i} + R_{i}(\phi
)
\]
for some polynomials $Q_{i}$ and $R_{i}$, so that
\[
\Tr P_{i} = {1 \over i}\Tr ( d Q_{i} + R_{i}(\phi ) )
\]
by the cyclic property of trace.

Then summing over $i$ gives the result.

\subsection{Extension to the full supercharge}
\label{sec:final}
The task now is to extend the result from $d$ to $\delta$. The
commutator terms in the definition of $\delta$ make
it much harder to deal with the trace. Specifically, $\delta$ does not
commute with the sum over cyclic permutations. Instead,
we proceed 
with a less direct approach, and make use of the result
for $d$.

Begin with a technical result:

\begin{it}
Suppose $P_{k}$ is a polynomial of degree $k$ satisfying 
$d \Delta \Tr P_{k}=0$. Then there exists a polynomial $Q$ such that
$d \Tr P_{k} = (d + \Delta ) \Tr Q$, as long as $N> {3k \over 2}$.
\end{it}

The proof follows an inductive argument. By (\ref{eq:dtrace}),
\[
d \Delta \Tr P_{k} = 0 \;\;\;\; \Rightarrow \;\;\;\; \Delta \Tr P_{k}
= - d \Tr P_{k+1} + R_{k+1}(\phi)
\]
for some polynomials $P_{k+1}$ and $R_{k+1}$ of degree $k+1$. Since
neither $d$ nor $\Delta$ can produce monomials only in $\phi$,
$R_{k+1}(\phi ) = 0$. Then
\[
d \Tr P_{k} = (d + \Delta ) \Tr P_{k} + d \Tr P_{k+1}
\]
On any monomial, $d$ acts to increase the number of fields of type
$B^{i}$ by $1$, 
whilst $\Delta$ acts to decrease the number of $B^{i}$ by $1$.

Let $M_{k}$ be the maximum number of $B^{i}$ occurring in any term of
$P_{k}$. Then since $\Delta \Tr P_{k} = d \Tr P_{k+1}$, we have
\[
M_{k+1} = M_{k} - 2
\]
Proceed inductively to find
\[
d \Tr P_{k} = (d+\Delta ) \Tr ( P_{k} + P_{k+1} + \cdots + P_{k+q})
+ d \Tr P_{k+q+1}
\]
where $P_{k+q+1}$ contains no $B^{i}$ fields at all. Then $\Delta \Tr
P_{k+q+1} = 0$ and so
\[
d \Tr P_{k} = (d + \Delta ) \Tr ( P_{k} + \cdots + P_{k+q+1} )
\]
which proves the result as long as $N>k+q+1$ so that each inductive
step is valid. Noting that the case $M_{k}=k$ is special and can be
reduced to the case $M_{k}=k-1$, one finds that $N>{3k \over 2}$ is a
sufficient condition.

We are now ready to prove the main result.

Suppose the matrix fields are of size $N$, and $P$ is a polynomial in
the matrix fields. Then for $\degree(P) < {2N \over 3}$,
\beq
\label{eq:cohom}
\delta \Tr P = 0 \Leftrightarrow \Tr P = \delta \Tr Q + \Tr R(\phi)
\eeq
where $Q$ and $R$ are polynomials.

To show this,
write $P= P_{0}+\cdots +P_{M}$ where $P_{i}$ contains monomials only
of degree $i$. Then
\[
\delta \Tr P = 0 \;\;\;\; \Rightarrow \;\;\;\; (d+\Delta ) \Tr P = 0
\]
and since $d$ preserves degree whilst $\Delta$ increases degree by
$1$, we have
\[
\begin{array}{l}
\Delta \Tr P_{M} = 0 \\
\Delta \Tr P_{i} + d \Tr P_{i+1} = 0,\;\;\;\; i=0,\cdots ,M-1 \\
d \Tr P_{0} = 0
\end{array}
\]
By (\ref{eq:dtrace}),
\[
d \Tr P_{0} = 0 \;\;\;\; \Rightarrow \;\;\;\; \Tr P_{0} = d \Tr Q_{0}
+ \Tr R_{0}(\phi )
\]
and
\[
\begin{array}{ccl}
\Delta \Tr P_{0} + d \Tr P_{1} = 0 & \Rightarrow & \Delta ( d \Tr
Q_{0} + \Tr R_{0}(\phi ) ) + d \Tr P_{1} = 0 \\
 & \Rightarrow & d ( - \Delta \Tr Q_{0} + \Tr P_{1} ) = 0 \\
 & \Rightarrow & \Tr P_{1} = \Delta \Tr Q_{0} + d \Tr Q_{1} + \Tr
R_{1} (\phi )
\end{array}
\]
By induction,
\[
\Tr P_{i} = \Delta \Tr Q_{i-1} + d \Tr Q_{i} + \Tr R_{i}(\phi
),\;\;\;\; i=1,\cdots ,M
\]
for some polynomials $Q_{i}$ and $R_{i}(\phi )$. So
\[
\begin{array}{ccl}
\Tr P &=& d \Tr Q_{0} + \Tr R_{0}(\phi) + \sum_{i=1}^{i=M} \Delta \Tr
Q_{i-1} + d \Tr Q_{i} + \Tr R_{i}(\phi) \\
&=& (d + \Delta ) \sum_{i=0}^{M-1} \Tr Q_{i} + \sum_{i=0}^{M} \Tr
R_{i}(\phi) + d \Tr Q_{M}
\end{array}
\]
Then
\[
\delta \Tr P = 0 \;\;\;\; \Rightarrow \;\;\;\; \Delta d \Tr Q_{M} = 0
\]
and so the technical result at the beginning of section \ref{sec:final} gives
\[
d \Tr Q_{M} = (d + \Delta) \Tr S
\]
for some polynomial $S$, and proves the result. 

\section{Concluding Remarks}
\setcounter{equation}{0}
We have considered the $SU(N)$ cohomological matrix models in four, six and
ten dimensions, and shown that
\beq
\label{eq:gen}
\delta \Tr P = 0 \Leftrightarrow \Tr P = \delta \Tr Q + \Tr R( \phi)
\eeq
as long as the  degree of $P$ is less than ${2N \over 3}$.

Although the large $N$ limit is a case of particular interest, it
would also be interesting to understand what happens when $N$ is
small, or the gauge group is not $SU(N)$. At present, we do not know of any
counter examples to the general formula (\ref{eq:gen}) in these cases.
It would also be interesting to understand whether the result
can be extended to a general gauge invariant quantity consisting of
an arbitrary function of traces. 

Finally we note that, in practice, the supercharge 
(\ref{eq:delta}) is often 
deformed $\delta \rightarrow \delta_{\rm{def}}$ to obtain useful results from
cohomological models 
\cite{src:kazakov, Hoppe:1999xg, Moore:1998et}. A typical
deformation leaves all field transformations under $\delta$ unchanged
except
\beq
\delta_{\rm{def}} \, \psi_{a}  =  [\phi,X_{a}] + i \nu \, \epsilon_{ab} X_{b}
\eeq
where $\nu$ is a parameter. We point out here that it is very easy to
adapt the methods of the previous section to obtain
\beq
\delta_{\rm{def}} \, \Tr P = 0 \Rightarrow \Tr P =
\delta_{\rm{def}} \, 
\Tr Q + \Tr R( \phi) 
\eeq
for $N$ sufficiently larger than $\degree(P)$.

\vspace{0.5 truecm}
\textbf{Acknowledgments} 

I am grateful to John Wheater for many valuable discussions. This
research is supported by PPARC.


\end{document}